\documentstyle{mn}
\input epsf
\newif\ifAMStwofonts

\def\lesssim{\mathrel{\hbox{\rlap{\hbox{\lower4pt\hbox{$\sim$}}}\hbox{$<$}}}}

\def\gtrsim{\mathrel{\hbox{\rlap{\hbox{\lower4pt\hbox{$\sim$}}}\hbox{$>$}}}}

\def\ll_lsun{$\log{L/\rm L_{\odot}}$~}

\def\masa_msun{$M/ \rm M_{\odot}$~}

\def\m_mstar{$M/M_{*}$~}

\def\msol{\,{\rm M}_{\odot}}

\title[Apsidal motion in HD\,93205]
{Calculation of the Masses of the  Binary Star  HD\,93205  by  
Application  of  the Theory of the Apsidal Motion}

\author[O.G.  Benvenuto, A.M. Serenelli,  L.G. Althaus,  R.H. Barb\'a,
N.I.  Morrell]  
{O.G.   Benvenuto\thanks{Member of the Carrera del Investigador 
Cient\'{\i}fico, CICBA, Argentina.  
Email: obenvenuto@fcaglp.fcaglp.unlp.edu.ar}, 
A.M. Serenelli\thanks{Fellow of CONICET,  Argentina.   
Email:  serenell@fcaglp.fcaglp.unlp.edu.ar},
L.G.   Althaus\thanks{Member of Carrera del Investigador Cient\'{\i}fico,
CONICET, Argentina}, R.H.  Barb\'a\raisebox{1ex}{$\ddagger$}, N.I.  Morrell\raisebox{1ex}{$\ddagger$}  \\  Facultad   de  Ciencias   Astron\'omicas  y
Geof\'{\i}sicas, Universidad  Nacional de  La Plata, Paseo  del Bosque
S/N, (1900) La Plata, Argentina}

\date{October 16}

\pagerange{\pageref{firstpage}--\pageref{lastpage}}

\pubyear{2001}

\begin{document}

\maketitle

\label{firstpage}

\begin{abstract}

We present a method to calculate masses  for components of
both  eclipsing  and non-eclipsing  binary  systems  as  long as  their
apsidal motion rates  are available.  The method is  based on the fact
that the  equation that gives  the rate of  apsidal motion is a 
supplementary equation  that allows the
computation of the masses of the components, if the radii and
the internal structure constants of them can be obtained from
theoretical models. For  this reason the use of this  equation makes the method
presented here {\it model dependent}.

We apply  this method to calculate  the mass of the  components of the
{\it  non-eclipsing} massive binary  system  HD~93205 (O3~V+O8~V), which is
suspected to be a very young system. To this end, we  computed a  grid  of 
evolutionary
models covering the mass range of interest, and
taking the  mass  of  the primary  $(M_1)$  as the  only
independent  variable, we  solve the  equation of
apsidal motion for $M_1$  as a  function of the age  of the
system. 
The mass of the primary we find  ranges from $M_1= 60\pm19\msol$ for ZAMS
models, which sets an upper limit for $M_1$, down to $M_1=  40\pm9\msol$ for 
an age of 2 Myr. Accordingly,  the upper limit derived for  the mass of the
secondary ($M_2= Q  M_1$) is $M_2= 25\msol$ is in very good
agreement with  the masses derived  for other O8~V stars occurring
 in eclipsing binaries. 

\end{abstract}


\section{Introduction} \label{sec:intro}


The motion of  the apse of a binary is mainly  a direct consequence of
the  finite size  of  its  components. If  both  stars were  spherical
objects and General Relativity corrections were negligible, they would
move  on a  Keplerian,  fixed  orbit.  However,  the  presence of  the
companion object, and  also its rotation, makes the  structure of each
star depart from spheres. In such a situation, there appears a finite
quadrupolar (and  higher) momentum to the gravitational  field of each
object that forces the orbit to  modify the position of the apse. This
is an effect well known from long time ago (Cowling 1938; Sterne 1939).
The rate of motion of the  apse is dependent on the internal structure
of  each  component;  thus, if  we  are  able  to determine  the  main
characteristics of a binary, it  provides an observational test of the
theory of stellar structure and evolution (Schwarzschild 1958; Kopal 1959).


In spite  of the age  of the  idea, apsidal motion  of binary
systems has been  systematically studied only recently in  a series of
papers by Claret \& Gim\'enez (e.g. Claret \& Gim\'enez 1993) and
Claret  (e.g. Claret  1995, 1997, 1998,  1999).   Perhaps, one  of the  main
reasons for such a situation is that the rate of motion of the apse is
very dependent on the stellar structure. Thus, the apsidal motion test
has been useful only recently  because of the availability of accurate
stellar  evolutionary  models  such as  those  of 
Claret (1995). These authors have performed detailed stellar 
models computing the  coefficients that determine  the rate of motion  of the
apse and  applied them to  compare with observational data from 
eclipsing binaries.

The detection of apsidal motion  in non-eclipsing binary systems is an
elusive subject. It has to be determined through the time variation of the
shape of the radial velocity curve  due to change in the
longitude of  periastron.  Generally,   the observed radial
velocities in binary systems  have large uncertainties that mask
this effect  in many  cases.  Moreover, the  fact that times  of light
minima  are  usually  determined  with high  precision  for  eclipsing
binaries, acts as a selection  effect in favour of detecting the motion
of  the line  of apsides  in such  systems: a  hundred cycles  are usually
enough  to notice  the change  in relative  position of  the secondary
minimum with respect to the primary minimum.  Observations over much longer
periods  of time are  needed to  find evidences  of apsidal  motion in
systems where eclipses are not  seen and the only observable effect is
the change in shape of the radial velocity orbit.

Empirical determinations of masses  are scarce for early O-type stars,
(e.g. Burkholder  et al. 1997;  Sch\"onberner \& Harmanec  1995).  Few
early O-type stars  are known to be members  of double-lined binaries,
and from them, those showing eclipses or some kind of light variations
that  enable  the  estimate  of  the orbital  inclination  (and  then,
absolute masses), are rare.  The so-called ``Mass Discrepancy'', first
described by  Herrero et al.   (1992) and recently  reviewed (Herrero,
Puls,  \& Villamariz  2000),  relates to  the  difference between  the
masses derived  via numerical  evolutionary models and  those obtained
from  spectral  analysis  (plus  model  atmospheres)  or  binary  star
studies.   This  discrepancy, amounting  to  50\%  in  1992, has  been
partially solved with the use of new evolutionary models that consider
the effect of  stellar rotation (Meynet \& Maeder  2000) and new model
atmospheres.  But large differences between `predicted' and `observed'
masses are  still present for  the hottest and youngest  (non evolved)
stars.   A recent  study  of the  massive  double-lined O-type  binary
system  HD\,93205 (Morrell  et  al.  2001)  yields  minimum masses  of
$31.5\pm1.1$\,M$_{\odot}$ and  $13.3\pm1.1$\,M$_{\odot}$ for the O3\,V
primary and O8\,V secondary components, respectively. This leads to
a probable  mass of  $\sim$52-60$M_{\odot}$ for the  O3\,V star,  if a
mass value according to those derived  for other O8\,V stars in eclipsing 
binaries
is assumed for  the secondary O8\,V. This is much less  than the 80 to
100  M$_{\odot}$  predicted  from  the  position of  this  star  on  a
theoretical HRD  compared to stellar evolutionary  tracks (see Fig.~7 in
Morrell et al. 2001).

Notably, HD~93205 is the first early O-type non-eclipsing
\footnote{Phase  dependent  light variations  with  full amplitude  of
$\sim 0.02$\,mag  in visual  light were reported  by Antokhina  et al.
(2000). These  authors stated that  the observed light  variations are
probably related to tidal  distortions rather than eclipses.} massive
binary for  which the rate of  motion of the apse  has been determined
with some  accuracy. Morrell et  al. (2001) derived an  apsidal motion
period  of $185\pm16$~years,  that considering  the orbital  period of
$6.0803\pm0.0004$    days,    yields    an   apsidal    motion    rate
$\dot{\varpi}=0\fdg0324\pm0\fdg0031$ per orbital  cycle.  This is very
interesting because,  from a mathematical  point of view, the  rate of
motion  of the apse  provides another  equation to  be applied  to the
system apart from the standard ones. HD~93205 is an early-type massive
short period  binary in a highly  eccentric orbit ($e=0.370\pm0.005$;
Morrell et al. 2001),
lying  in the  Carina Nebula,  a galactic  massive star forming
region.  Thus, we can assume its  components are on, or very close to,
the Zero  Age Main-Sequence (ZAMS).  Consequently, if  we consider the
evolutionary stage of HD~93205 as  known, the equation of apsidal
motion can be  written in a way that  we can solve it for  the mass of
the primary  star.  The  aim of the  present paper  is to
detail this method and to apply it to HD~93205.

The paper is organized  as follows: in Section
\ref{sec:compu} we present  the method we use to  obtain the masses for
 components  of non-eclipsing  binary  systems and we describe  our
evolutionary code and the calculations we have carried out. In Section
\ref{sec:preli} we present a test of our method by applying it to some
eclipsing binary systems.  Section  \ref{sec:soluc} is devoted to showing
the  results  obtained for  the  massive  binary  system HD~93205  and
finally, in  Section \ref{sec:conclu} we give  some concluding remarks
about the implications of our results.

\section{Computational details} \label{sec:compu}

Here  we present an original method, as  far as the authors
are aware, to calculate the masses of the components of binary systems
provided  the knowledge  of the  rate of  motion of  the apse  and the
evolutionary status of  the stars.  In addition, we  describe the main
characteristics of  our evolutionary code  and the calculation  of the
internal structure  constants (ISC) on  which the rate of  the apsidal
motion depends.

\subsection{Equations of apsidal motion and description of the method} 
\label{subsec:equations}

Sterne (1939) has shown  that if the classical gravitational potential
of  each component  of a  binary  system is  expanded in  a series  of
spherical harmonics, and terms  up to the quadrupolar contribution are
kept, then the rate of motion of the apse is given by \footnote{It is
assumed that rotation of both components is perpendicular to the orbital
plane.}
\begin{eqnarray}
\label{eq:sterne}
\frac{\dot{\varpi}_{2}}{\Omega}= k_{2,1} \left(\frac{a_{1}}{A}\right)^{5}
\left[   15   \frac{M_{2}}{M_{1}}   f_{2}(e)  +   \frac{\omega^{2}_{1}
A^{3}}{M_{1}   G}   g_{2}(e)    \right]   +   \\   \nonumber   k_{2,2}
\left(\frac{a_{2}}{A}\right)^{5}    \left[    15   \frac{M_{1}}{M_{2}}
f_{2}(e) + \frac{\omega^{2}_{2} A^{3}}{M_{2} G} g_{2}(e) \right],
\end{eqnarray}
where $\dot{\varpi}_{2}$ is the rate of the secular motion of the apse
calculated considering  only the quadrupolar  contribution (the lowest
order)  of the gravitational  potential; $k_{i,j}$  are the  ISCs that
depend  on the  internal mass  distribution of  the stars  (see below,
Subsection \ref{subsec:isc} for more  details); $i$ denotes the $i$-th
multipolar  momentum considered  (i.e.  $i=2$  throughout  this paper);
whereas $j$  denotes the component of  the binary system.   $G$ is the
gravitational constant,  $\Omega$ is  the mean orbital  angular velocity,
$A$ is the semiaxis of the relative orbit, $a_{1}$ and $a_{2}$ are the
mean  radii of  the  stars, $\omega_{1}$  and  $\omega_{2}$ are  their
angular velocities  of rotation, $M_{1}$ and  $M_{2}$ are
the  stellar masses; and finally $f_{2}(e)$  and $g_{2}(e)$  are
functions of the orbital eccentricity $e$ given by
\begin{equation}
f_{2}(e)=  \left( 1 +  \frac{3}{2} e^{2}  + \frac{1}{8}  e^{4} \right)
\left( 1 - e^{2} \right)^{-5},
\end{equation}
\begin{equation}
g_{2}(e)= \left( 1 - e^{2} \right)^{-2}.
\end{equation}
In the  following we shall reformulate  Eq.~(\ref{eq:sterne}) in order
to write it to become an implicit equation for the mass of the primary
star  $M_{1}$. The semiaxis $A$ is not directly known from observations, 
however the projected semiaxis $D$ given by

\begin{equation}
D= A \sin{i} 
\end{equation}

\noindent where $\sin{i}$ is the sine of the inclination of the orbit,
can be observationally assessed. Let  us  define the  mass  ratio $Q$  
and the  angular velocities ratio $q_{\omega}$ as

\begin{equation}
Q= \frac{M_{2}}{M_{1}} ; \;\;\; q_{\omega}= \frac{\omega_{2}}{\omega_{1}}.
\end{equation}
%
%
%
%
%
%
In order to eliminate $\sin{i}$ we can use the mass function, defined as
\begin{equation}
f=  \frac{  M^{3}_{1}  \sin^{3}{i}   }  {  \left(  M_{1}  +  M_{2}
\right)^{2} }= \frac{  M_{1} \sin^{3}{i} } { \left(  1 + Q \right)^{2}
},
\label{eq:fmasa}
\end{equation}
which can be determined from observations (Batten 1973). In the same trend, 
projected tangential velocities $v_1=V_1 \sin{i}$ and $v_2=V_2 \sin{i}$ are 
also observable quantities and their ratio $q=v_2/v_1$ can be used to 
eliminate $q_{\omega}$.

One important  point  is that  rotation  modifies the  internal
structure of  the stars.  In a  recent paper, Claret  (1999) has shown
that, within the quasi-spherical  approximation, rotation can be taken
into account in the apsidal motion analysis simply by reducing the ISC
$k_{2,i}$ by

\begin{equation}
\log{k_{2,i}}= \log{[k_{2,i}]_{sph}} - 0.87 \lambda_i.
\label{eq:rot}
\end{equation}

\noindent  Here,  $[k_{2,i}]_{sph}$  denotes  the  ISC  obtained  from
spherical models and the parameter $\lambda_i$ is defined by
\begin{equation}
\lambda_i = \frac{2V^{2}_{i}}{3 g_{i} a_{i}}
\end{equation}

\noindent where  $i$ denotes  the component of  the binary  system and
$V_i$, $a_i$ and $g_i$ are, respectively, the tangential velocity, the
radius and the surface gravity of the component.

Up to  this point,  we have only  considered the contributions  to the
motion of the apse due to Newtonian gravity. However, it is well known
that General Relativity predicts a secular motion of the apse which is
independent of  the classical  contributions. The angular  velocity of
the apse due to General Relativistic effects $\dot{\varpi}_{GR}$ is given by
(Levi-Civita 1937)

\begin{equation}
\frac{\dot{\varpi}_{GR}}{\Omega}=              6.36\times             10^{-6}\;
\frac{M_{1}+M_{2}}{A\left(1-e^{2}\right)}
\label{eq:grc1}
\end{equation}
 
Defining ${\cal F}_{2}$ as 
\begin{equation}
{\cal F}_{2}= \frac{\dot{\varpi}}{\Omega}  - 
\left( \frac{\dot{\varpi}_{GR}}{\Omega} + \frac{\dot{\varpi}_{2}}  {\Omega} 
\right) = 0
\end{equation}
and incorporating both, rotation and relativistic effects we get after some 
algebraic manipulation
\begin{eqnarray}
{\cal F}_{2}= \frac{\dot{\varpi}}{\Omega}  - \left\{ 6.36\times 10^{-6}\;
\frac{\left[   fM_{1}^2(1+Q)^5\right]^{1/3}}{D\left(1-e^{2}\right)}
+ \  \ \  \ \ \  \ \ \  \ \  \ \nonumber \right.   \\ \frac{15}{D^{5}}
f_{2}(e) \left[  \frac{ f  \left( 1  + Q \right)^{2}  } {  M_{1} }
\right]^{5/3}   \left[  k_{2,1}  a^{5}_{1}   Q  +   k_{2,2}  a^{5}_{2}
\frac{1}{Q}   \right]  +   \label{eq:fundam}   \\  \left.    \nonumber
\frac{v^{2}_{1} }{M_{1}  G D^{2}} g_{2}(e) \left[  k_{2,1} a^{3}_{1} +
k_{2,2} a^{3}_{2} \frac{q^{2}}{Q} \right] \right\} = 0 \ \ \ \ \ \ \ \
\ \ \ \ \ \
\end{eqnarray}

\noindent where  $k_{2,i}$ are  the ISCs corrected  by the  effects of
rotation. This is the fundamental equation for our purposes.

As  mentioned  above, in  Eq.~(\ref{eq:fundam}) some  quantities  are
determined observationally  ($\Omega, \dot{\varpi}_2, e, D,  Q, v_1, f,
q$). On the other hand, $k_{2,1}$ and $k_{2,2}$ must be computed from
evolutionary models and, if we are dealing with non-eclipsing systems, 
as is the case for HD~93205, then the radii $a_{1}$ and $a_{2}$ must be
obtained from theoretical models as well. 
Now, if  we assume that  both components  of the
binary  system  have  the  same  age, and  we  use  the  observational
constraint $Q= M_2 / M_1$,  then $k_{2,1}, k_{2,2}, a_{1}$ and $a_{2}$
can be  derived from evolutionary calculations as a  function of
$M_1$ and the age of the system.

The  method presented  here  to  calculate $M_1$  is  as follows:  we
compute a grid of evolutionary models covering the range of
masses  of interest  with  a small  mass  step.  Using  this grid,  we
construct isochrones starting at the  ZAMS with a given time step and
for each isochrone we seek  the solution of Eq.~(\ref{eq:fundam}).  In 
this  procedure, the only independent quantity is
$M_1$  so when  the solution  of Eq.~(\ref{eq:fundam})  is  found, the
corresponding  value of $M_1$  is the  mass of  the primary  star that
corresponds to the age of the isochrone.

Thus, for a given age of the  system we have one solution for the mass
of the primary star $M_1$, and using  $Q= M_2 / M_1$ we can derive the
mass of the secondary. However, the age  of  the  system  must  be  
constrained by other means. In  addition, the  masses of the  components 
that  can be
found  with the  present  method are  {\it  model dependent}.   Notice
especially  the  sensitivity of  the  tidal  and  rotational terms  in
Eq.~(\ref{eq:fundam}) to the value of  the stellar radius. It is clear
that we need accurate stellar models in order to get
a physically  reliable value  of  $M_{1}$.

Finally,   it    is   worth   mentioning   that    the   solution   of
Eq.~(\ref{eq:fundam}) is subject to some constraints. Two of them were
already mentioned: the age  of both components  in a binary
system must be the same, and $M_2= Q\,M_1$. In addition, the value
of the mass function imposes a minimum value for $M_{1}$

\begin{equation}
\left[M_{1}\right]_{min}= f\; \left( 1 + Q \right)^{2}.
\end{equation}

\noindent  and also,  the  condition  for the  system to  be
detached
\begin{equation}
a_{1} + a_{2} < D,
\end{equation}
\noindent must be fulfilled.

\subsection{Evolutionary models and calculation of the ISCs} 
\label{subsec:isc}

As stated above, in order  to solve Eq.~(\ref{eq:fundam}), the ISCs and
radii  of   both  components   must  be  obtained   from  evolutionary
calculations.   This leads  us to  the necessity  of having  a  set of
evolutionary tracks  of objects covering the range  of masses expected
for the components of the system. In the case of HD~93205, the primary
O3\,V star  is candidate to  be one of  the most massive  stars known.
Thus,  we have  carried out  calculations  up to  quite large  stellar
masses such  as 106~$\rm M_{\odot}$ because,  as far as  we are aware,
there is no  computation of the ISCs for  such massive stars available
in the literature.

The calculations have been carried out with the stellar evolution code
developed at  La Plata  Observatory. It is  essentially the  same code
 employed for  studying white dwarf  (see, e.g., Benvenuto
\&  Althaus 1998) and  intermediate mass  stars (Brunini  \& Benvenuto
1997) and has  been adapted  for properly handling the  case of  massive 
stars.

Let  us  briefly describe  the  main  ingredients  of the code.   The
equation  of state employed is  that of  OPAL (Rogers,  Swenson \&
Iglesias  1996). Radiative opacities  are the  latest version  of OPAL
(Iglesias  \&  Rogers  1996)  while  for  low  temperatures  they  are
complemented   with  the  Alexander   \&  Ferguson   (1994)  molecular
opacities. Conductive opacities and  neutrino emission rates
are the same as
in Benvenuto \& Althaus (1998). Nuclear  reaction rates are  taken from  
Caughlan \& Fowler (1988)  and weak electron  screening is taken from  
Graboske et al. (1973). 

As we are dealing with massive  stars, it is important to mention that
we have accounted for the  occurrence of overshooting by employing the
formalism described  in Maeder \&  Meynet (1989).   
We have  adopted the  distance of  overshooting $d_{ov}$ to  be a  fraction  
of  the pressure  scale  height  $H_P$ at  the
canonical border of the  convective zone: $d_{ov}= \alpha_{ov} H_{p}$.
Also, we allowed for mass loss following De Jager et al. (1986).

Using the evolutionary  code just described, we have  calculated a set
of  evolutionary  sequences  covering  the  mass  range  from  $4  \rm
M_{\odot}$  to $106  \rm M_{\odot}$ with  a mass step  of $\approx$
5\%.   We  followed the  evolution  starting at  the  ZAMS  till  the
depletion of hydrogen at the centre of the star. The initial helium
content of
our models is $Y=0.275$ and the adopted value for
metallicity   is   $Z=0.02$  while   two   values  for   overshooting,
$\alpha_{ov}= 0.25$ and $0.40$, were considered.

After  convergence   of  each  model   was  reached,  we   solved  the
Clairaut-Radau differential  equation (Sterne 1939)  that accounts for
the apsidal motion to the lowest (second) order

\begin{equation}
a   \frac{d\eta_{2}}{da}  +   \eta^{2}_{2}  -   \eta_{2}  -   6   +  6
\frac{\rho}{\bar{\rho}} \left( \eta_{2} + 1 \right) = 0
\label{eq:radau}
\end{equation}
subject to the boundary condition $\eta_2= 0$ at $a=0$.
In this expression $a$  is the  mean radius of a given 
equipotential, $\rho(a)$  is the density  at $a$  and $\bar{\rho}(a)$  
is the mean density interior to $a$, $\eta_i$ is given by
\begin{equation}
\eta_i \equiv \frac{a}{Y_i} \frac{dY_i}{da}
\end{equation}
and the radius  $r$ of the distorted configuration and $a$ are related by,
\begin{equation}
r=a\left( 1+ \sum_{i=0}^{n} Y_i(a,\theta) \right)
\end{equation}
where $Y_i(a,\theta)$ describe the amplitude of the distortions.
In order to integrate Eq.~(\ref{eq:radau}) we recall that $\rho(a)$ and 
$\bar{\rho}(a)$ are provided by the structure of the evolving model.
Close to the centre $\rho(a)$, and consequently $\bar{\rho}(a)$ and
$\eta_2(a)$, are expanded following an analogous treatment to that 
presented in Brooker \& Olle (1955). The integration of Eq.(\ref{eq:radau})
is started at the mesh point adjacent to the centre. 
Numerical integration is carried 
out with a standard Runge-Kutta routine (Press  et al.  1986) up to  the 
surface of the  stellar model in order to get $\eta_2(a_i)$. Then, the ISC 
$k_{2,i}$ is finally given by
\begin{equation}
k_{2,i}=          \left[          \frac{3-\eta_{2}(a)}{4+2\eta_{2}(a)}
\right]_{a=a_{i}}.
\end{equation}


\section{A test of the method employing eclipsing binary systems}
\label{sec:preli}

The method we  are presenting here is, to  our knowledge, original. In
view  of this fact,  we have  applied this  method to  some previously
studied massive eclipsing binary  systems with the aim of testing the method
before applying it to HD 93205. 
We  have focused our
attention  on detached systems  in which  both components  are massive
stars in the  main sequence (MS) and have a rather  well measured apsidal 
motion.
We  have  finally  selected  the following  systems:  EM~Car,  QX~Car,
GL~Car, Y~Cyg and V478~Cyg. Observational parameters for these systems
are summarized  in Table~\ref{tab:gen}.  In order to test the method, we  
compare  the masses it yields with the  {\em observed} ones, i.e. those 
obtained
from the simultaneous analysis of light and radial velocity curves of
the systems.

\begin{figure*}
\epsfxsize=400pt
\epsfbox[19 320 578 800]{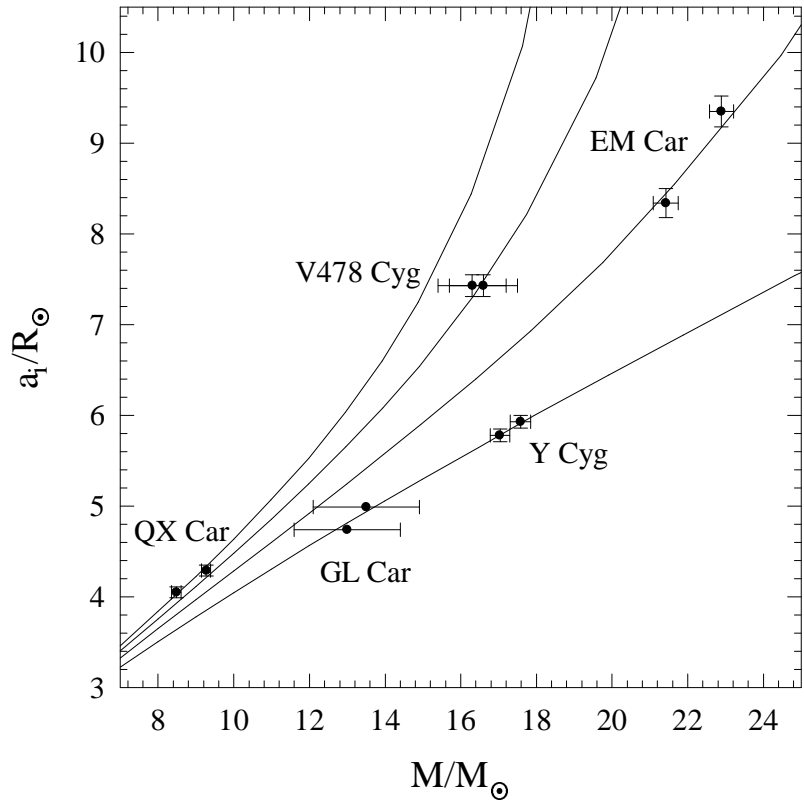} 
\caption{The radius vs. mass relationship for the components of binary
systems EM Car, QX Car, Gl Car, Y Cyg and V478 Cyg together with their
corresponding  error  bars.  Solid  lines  represent  our  theoretical
isochrones for 7.8, 6.3, 4.3 and 1.8 Myr (from right to left). For each 
system considered there is one isochrone that fits both components.
\label{fig:m-r}} 
\end{figure*}

\begin{figure*}
\epsfxsize=400pt
\epsfbox[19 320 578 800]{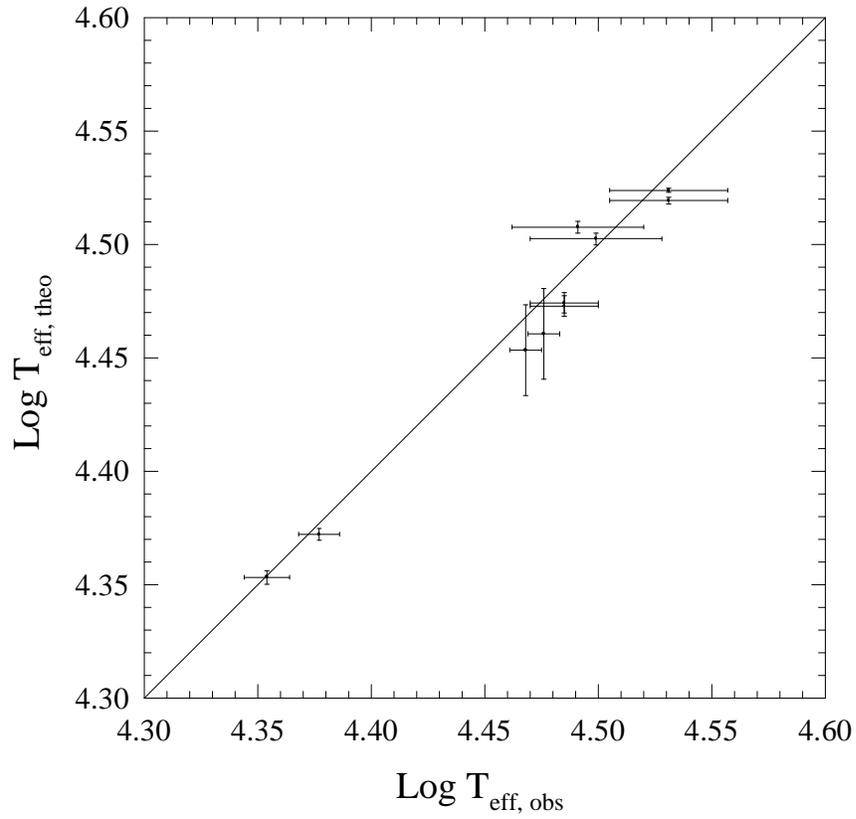} 
\caption{Comparison  between   theoretical  and  observed  effective
temperatures for the components of the same binary systems shown in 
Fig.~\ref{fig:m-r}.
\label{fig:teff}}
\end{figure*}

Components of  a binary system  must have the  same age, so  the first
test  we  apply to  the  evolutionary code  is  that  for each  system
considered  there must  be a single isochrone fitting the  mass  and 
radius  of both  components on  the $M-R$  plane. The  results  of our
calculations for  the choice $\alpha_{ov}=0.25$ can  be appreciated in
Fig. \ref{fig:m-r}, in  which we show the mass and  radius of the 
components of  the selected binary systems. Note that  for each system
there  is  one  isochrone  that  fits well  both  components,  so  the
constraint  that the  ages of  the components  impose  on evolutionary
calculations    is   clearly    satisfied   by    our    models.    In
Fig.~\ref{fig:teff}  we show the  effective temperatures  derived from
our  models for  each star  as a  function of  the  observed effective
temperature.   Again,  a   good  agreement  between  our  evolutionary
calculations and observations is found.  As we have already stated, we
have  also  considered a  larger  amount  of  overshooting, by  fixing
$\alpha_{ov}$  to 0.40. We  find no  significant differences  with the
case of a smaller amount of  overshooting, so we adopt the lower value
($\alpha_{ov}=0.25$) as the standard one in our calculations.

Based on these  preliminary results, we are confident  that our models
are appropriate for studying the evolution of stars in detached binary
systems.  Thus,  we apply  our models to  the study of  apsidal motion
through  the   calculation  of  the  ISCs.   Let   us  consider  again
Eq.~(\ref{eq:sterne}).  We  can rewrite it  in order to define  a {\em
mean} observational ISC $\overline{k}_{2,obs}$

\begin{eqnarray}
\frac{\dot{\varpi}_2}{\Omega}   =  k_{2,1}c_{2,1}   +   k_{2,2}c_{2,2}  =
\overline{k}_{2,obs}(c_{2,1} + c_{2,2}),
\end{eqnarray}
\noindent and a {\em mean} theoretical ISC by
\begin{eqnarray}
\overline{k}_{2,theo} = \frac{k_{2,1}c_{2,1}+ k_{2,2}c_{2,2}}{c_{2,1}+
c_{2,2}}.
\end{eqnarray}
\noindent where $c_{2,i}$ is given by
\begin{equation}
c_{2,i}= \left(\frac{a_{i}}{A}\right)^{5}
\left[   15   \frac{M_{3-i}}{M_{i}}   f_{2}(e)  +   \frac{\omega^{2}_{i}
A^{3}}{M_{i}   G}   g_{2}(e)    \right]. 
\end{equation}

\begin{figure*}
\epsfxsize=400pt
\epsfbox[19 320 578 800]{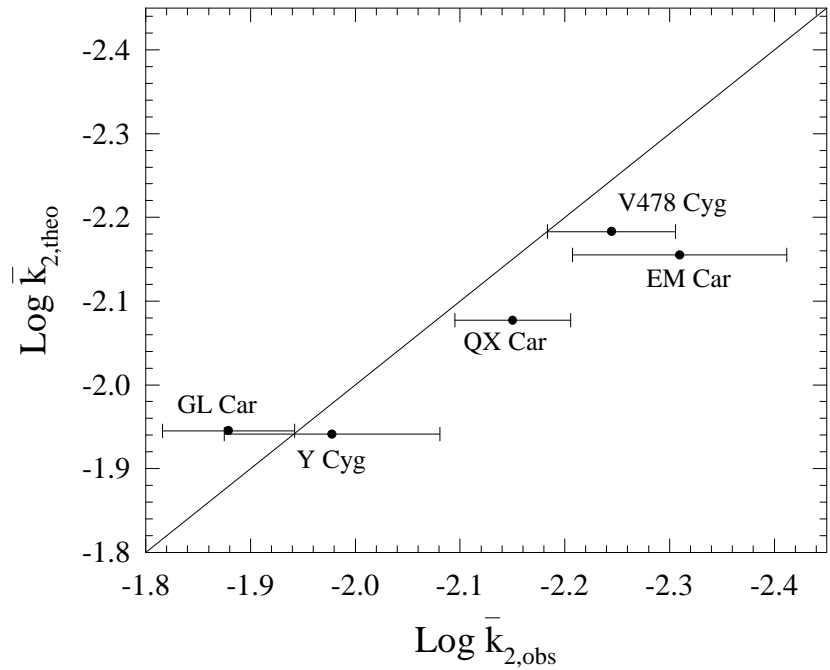} 
\caption{Comparison      between     theoretical      and     observed
$\overline{k}_{2}$ for  the components of the  same binary systems
included in Fig.~\ref{fig:m-r}. 1 $\sigma$ error bars for 
$\overline{k}_{2,obs}$ are also shown. Notice that theoretical models are 
slightly less concentrated than indicated by observations.  
\label{fig:isc}}
\end{figure*}

Even  when  the quotient  $\dot{\varpi}_2/\Omega$  can  be assessed  from
observation, it  is not possible  to separate the contribution  of each
component to the
rate of  apsidal motion.   Instead, we  can
determine $\overline{k}_{2,obs}$ and it is  this value that is
currently  used to  contrast evolutionary  models with  observation.  In
Fig.~\ref{fig:isc}    we    show    the   theoretical    values    for
$\overline{k}_{2}$ derived from our  models against the observed ones.
We find that our models predict  mean ISCs that are in reasonable good
agreement  with the  observed ones  for the  less  concentrated models
(those with a higher value of $\overline{k}_{2}$). The most discrepant
case  we find  is that  of EM~Car,  which is  the most  evolved system
considered by us, as we find that its primary star has spent $\approx$
60\%  of its life  on the  MS. In  this case,  our models  result less
concentrated  than  what we  should  expect  from  the observed  value
$\overline{k}_{2,obs}$. However, as stated by Andersen \& Clausen (1989),
the apsidal motion rate for this system is based on observations covering
only about 1/6 of the apsidal motion period, thus the accuracy of 
apsidal motion parameters is still limited. In addition, information on
its chemical composition is also missing. 
The theoretical values $\overline{k}_{2,theo}$ we find
for the other systems  are in good agreement with the observed values and, 
for QX~Car and Y~Cyg, also with those derived theoretically by Claret (1997).

\begin{figure*}
\epsfxsize=400pt
\epsfbox[19 320 578 800]{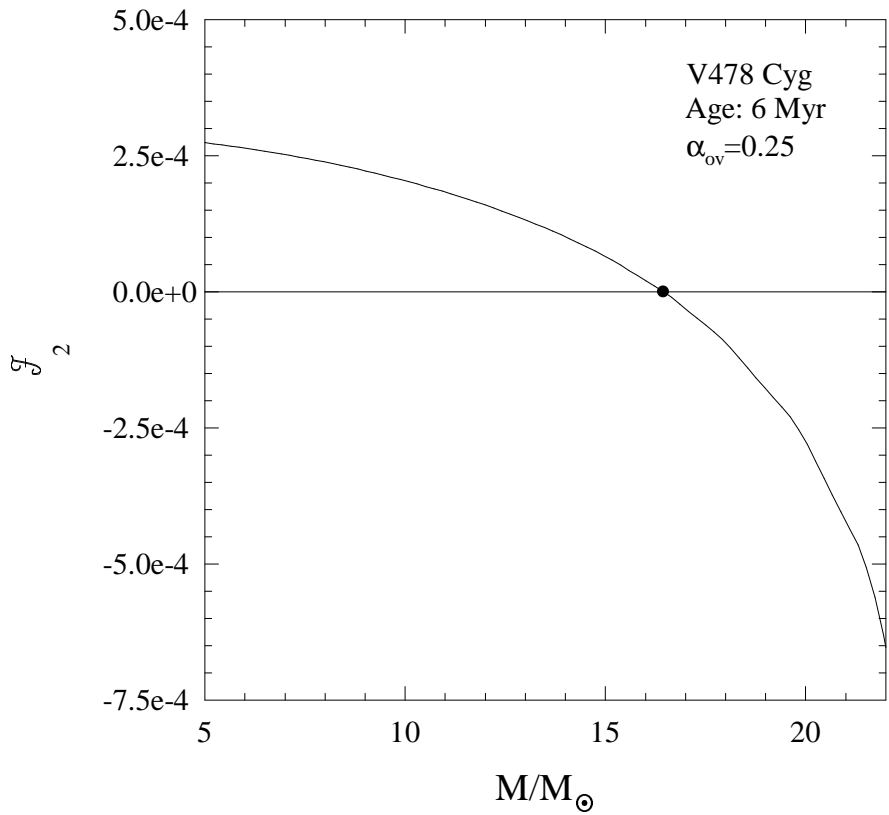} 
\caption{${\cal F}_2$ as a function of $M_1$ for the case of V478~Cyg,
assuming   an  age   of   6.0   Myr  for   the   system  and   setting
$\alpha_{ov}=0.25$.   Note that  the  solution value  $M_1$ for  which
${\cal  F}_2=0$ is  very  well  determined, since  ${\cal  F}_2$ is  a
monotonously decreasing function of $M_1$.
\label{fig:soluc}}
\end{figure*}

\begin{figure*}
\epsfxsize=400pt
\epsfbox[100 450 578 800]{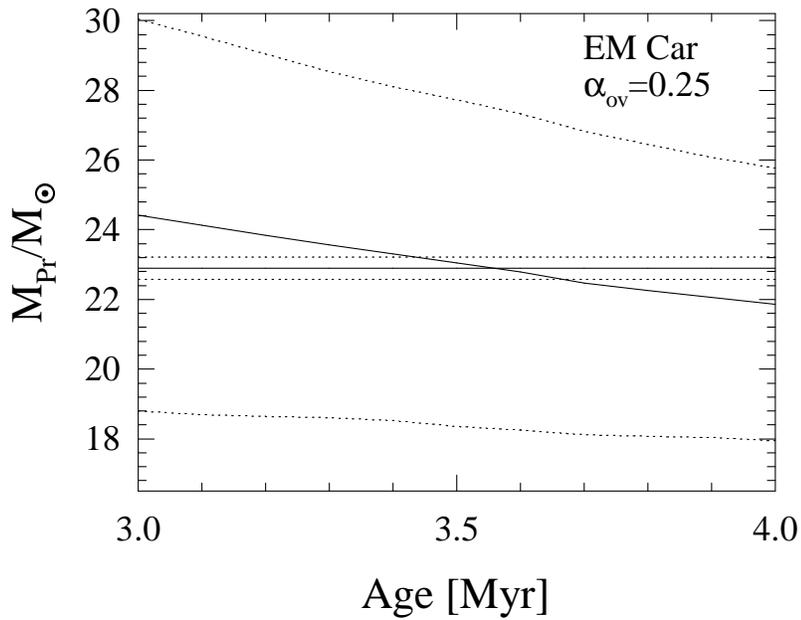} 
\caption{The mass  of the primary star  of EM  Car system assuming
$\alpha_{ov}=0.25$  as a function  of its  age. Horizontal  solid line
corresponds to  the preferred value  deduced from radial  velocity and
light curves. Short dashed  horizontal lines represent the uncertainty
in this  {\it observational value}.  The other solid and  short dashed
lines  represent the  preferred  value and  its $1\sigma$  uncertainty
respectively, deduced from the apsidal motion rate.
\label{fig:emcar25}}
\end{figure*}

\begin{figure*}
\epsfxsize=400pt
\epsfbox[50 450 578 800]{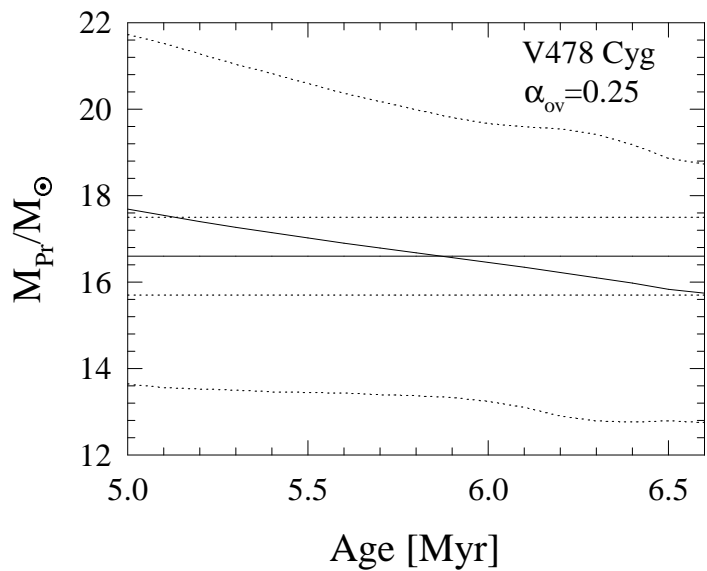} 
\caption{Same as  Fig.~\ref{fig:emcar25} but for V478 Cyg.
\label{fig:v478cyg25}}
\end{figure*}

We present  below some of  the results of  applying our method  to the
solution  of Eq.~(\ref{eq:fundam}).   First of  all, let  us emphasize
that  for   an  assumed  age   of  the  system,  the   solution  of
Eq.~(\ref{eq:fundam}) is very well  determined, i.e. only one solution
is  found  as  ${\cal  F}_2$  is  a very well-behaved,  monotonously
decreasing function  of the independent quantity  $M_1$. We illustrate
this general  behaviour with  one example: in  Fig.~\ref{fig:soluc} we
show ${\cal  F}_2$ as  a function  of $M_1$ for  the case  of V478~Cyg
assuming an age of 6 Myr for both components.  
In view of these results, we find the
method  to be very  reliable, from  a mathematical  point of  view, in
yielding a well determined  value of $M_1$. In Fig.~\ref{fig:emcar25}
it  is  shown  the mass  $M_1$  of the  primary component of EM~Car obtained 
as a function of the age of the system. The
figure  corresponds to  the choice $\alpha_{ov}=0.25$
for overshooting. We  find a  good
agreement between  our theoretical prediction  for $M_1$ and  its {\em
observed} value for the whole  range of ages considered, within a $\pm
1 \sigma$  error. The upper limit  for the age considered  is given by
the fact  that, for  larger ages, $M_1$  falls below the  minimum mass
derived   for   this   system    from   its   mass   function   $f$. 
Results are very similar if an overshooting amount of $\alpha_{ov}=0.40$
is considered, the only main difference being that  solution curve 
is slightly  shifted to  larger ages  ($\approx$  10\%). 
Fig.~\ref{fig:v478cyg25}  shows the  results obtained for V478~Cyg again
for $\alpha_{ov}=0.25$. For  this binary system, an excellent agreement
is achieved between our method  and the {\em observed} mass. 
The same trend  as before is  found, i.e.  the
larger  overshooting  shifts  the  solution  toward  ages  about  10\%
larger. Note also that the  larger the age considered, the smaller the
mass of  the primary  (and also  the mass of  the secondary)  that can
account for the observed rate of apsidal motion.
\begin{center}
\begin{table*}
\begin{minipage}{162mm}
\caption{Astrophysical   parameters    for   selected   test   systems
\label{tab:gen}}
\begin{tabular*}{162mm}{lccp{8pt}ccp{8pt}ccp{8pt}ccp{8pt}cc} \hline
& \multicolumn{2}{c}{EM Car} && \multicolumn{2}{c}{GL Car} && 
\multicolumn{2}{c}{QX Car} && \multicolumn{2}{c}{Y Cyg} && 
\multicolumn{2}{c}{V478 Cyg}\\ \hline
P [days] & \multicolumn{2}{c}{3.415} && \multicolumn{2}{c}{2.422} &&
\multicolumn{2}{c}{4.478} && \multicolumn{2}{c}{2.996} &&
\multicolumn{2}{c}{2.881} \\
$\dot{\varpi}$ [$^{\circ}$ day$^{-1}$] & \multicolumn{2}{c}{0.0237} && 
\multicolumn{2}{c}{0.03910} &&
\multicolumn{2}{c}{0.0027} && \multicolumn{2}{c}{0.0206} &&
\multicolumn{2}{c}{0.01301}\\
& \multicolumn{2}{c}{0.0029} && \multicolumn{2}{c}{0.00005} &&
\multicolumn{2}{c}{0.00005} && \multicolumn{2}{c}{0.00008} &&
\multicolumn{2}{c}{0.00134}\\
$e$ & \multicolumn{2}{c}{0.0120} && \multicolumn{2}{c}{0.1457} &&
\multicolumn{2}{c}{0.278} && \multicolumn{2}{c}{0.142} &&
\multicolumn{2}{c}{0.019}\\
& \multicolumn{2}{c}{0.0005} && \multicolumn{2}{c}{0.0010} &&
\multicolumn{2}{c}{0.003} && \multicolumn{2}{c}{0.002} &&
\multicolumn{2}{c}{0.002}\\
$A$ [R$_{\odot}$] & \multicolumn{2}{c}{33.70} && \multicolumn{2}{c}{22.64} &&
\multicolumn{2}{c}{29.79} && \multicolumn{2}{c}{28.44} &&
\multicolumn{2}{c}{27.32}\\
& \multicolumn{2}{c}{0.15} && \multicolumn{2}{c}{0.62} &&
\multicolumn{2}{c}{1.04} && \multicolumn{2}{c}{0.2} &&
\multicolumn{2}{c}{0.64}\\
$i$ [$^{\circ}$] & \multicolumn{2}{c}{81.5} && \multicolumn{2}{c}{86.4} &&
\multicolumn{2}{c}{85.7} && \multicolumn{2}{c}{85.5} && \multicolumn{2}{c}{78.0}\\
& \multicolumn{2}{c}{0.2} && \multicolumn{2}{c}{0.2} &&
\multicolumn{2}{c}{0.2} && \multicolumn{2}{c}{0.5} && \multicolumn{2}{c}{0.6}\\
\hline
& Prim. & Sec. && Prim. & Sec.&& Prim.& Sec. && Prim. & Sec. && Prim. & Sec.\\ 
\hline
Sp & O8V & O8V && B0.5 & B1 && B2V & B2V && O9.3 & O9.4 && O9.5V & O9.5V \\
log $T_{\rm eff}$ & 4.531 & 4.531 && 4.476 & 4.468 && 4.377 & 4.354 && 
4.491 & 4.499 && 4.485 & 4.485 \\
& 0.026 & 0.026 && 0.007 & 0.007 && 0.009 & 0.010 && 
0.029 & 0.029 && 0.015 & 0.015 \\
log $g$& 3.926 & 3.856 && 4.17 & 4.2 && 4.140 & 4.151 && 
4.12 & 4.17 && 3.916 & 3.908 \\
& 0.17 & 0.17 && - & - && 0.014 & 0.015 && 
0.04 & 0.04 && 0.027 & 0.027 \\
$M$ [M$_{\odot}$] & 22.89 & 21.42 && 13.5 & 13.0 && 9.27 & 8.48 && 17.57 & 17.04 
&& 16.6 & 16.3 \\
& 0.32 & 0.33 && 1.4 & 1.4 && 0.122 & 0.122 && 0.27 & 0.26 
&& 0.9 & 0.9 \\
$a$ [R$_{\odot}$] & 9.35 & 8.34 && 4.99 & 4.74 && 4.29 & 4.05 && 5.93 & 5.78 
&& 7.43 & 7.43 \\
& 0.17 & 0.16 && - & - && 0.06 & 0.06 && 0.07 & 0.07 
&& 0.12 & 0.12 \\
$V$ [km/s] & 150 & 130 && 141 & 134 && 120 & 110 && 147 & 138 && 135 & 135 \\ 
& 20 & 15 && SR & SR && 10 & 10 && 10 & 10 && SR & SR \\ 
Refs.& \multicolumn{2}{c}{1} && \multicolumn{2}{c}{2} &&
\multicolumn{2}{c}{3,4} && \multicolumn{2}{c}{5.6} &&
\multicolumn{2}{c}{7} \\ \hline
\end{tabular*}

{\small SR: sinchronous rotation is assumed.

Refs.: (1) Andersen \& Clausen (1989), (2) Gim\'enez \& Clausen (1986),
(3) Gim\'enez, Clausen \& Jensen (1986), (4) Andersen et al. (1983), 
(5) Simon, Sturm \& Fiedler (1994), (6) Hill \& Holmgren (1995),
(7) Petrova \& Orlov (1999) and references therein. }
\end{minipage}
\end{table*}
\end{center}\section{Calculation of the masses of HD~93205 and related parameters}
\label{sec:soluc}

From the results  of the previous sections, we judge  our method to be
good  enough to  be  employed in  the  mass estimation  of the  
components  of HD~93205.   As stated
before in the  Introduction, HD~93205  is  a  highly eccentric  system,
which strongly  suggests that it
must be very young.  However, the age estimates for such  early O-type
stars are very  uncertain, either one tries to  derive them considering
the  region in  which the  star is  located, or  comparing  the star's
position  on the  theoretical H-R  diagram with  isochrones calculated
from evolutionary stellar models. HD~93205 belongs to the open cluster
Trumpler 16, the  most  massive stars of which have 
 an age spread  between 1\,Myr  and 2\,Myr (DeGioia-Eastwood et al.  2001).
 Besides, there is
evidence  of ongoing  star formation  in the  molecular  cloud complex
associated with the Carina Nebula (Megeath et al. 1996). Consequently,
a  lower  limit  to  the  age  of the  members  of  Tr\,16  cannot  be
established.   On the  other hand,  de Koter,  Heap, \&  Hubeny (1998)
showed that if we increase by about 10\% the effective temperature of
O3-type stars, the age would decrease from 2\,Myr to
1\,Myr.  Regarding  the interpretation of  theoretical isochrones for
the  most massive  stars, these  authors stated: {\em  ``The derived
T$_{\rm eff}$  values are so  similar because the isochrone  for $\sim
2$\,Myr runs  almost vertical and because the  distance in temperature
between the isochrones of 1 and 3 Myr is very small''}.

Taking into account the problem  in the age determination described in
the previous paragraph we  choose to solve Eq.~(\ref{eq:fundam}) for a
whole set  of isochrones ranging  from the ZAMS up to 2 Myr. We  consider as
zero age isochrone the one  corresponding to the time when the stellar
radius reaches its minimum value. In our models, this happens for ages
of a few ten thousand years.

\begin{figure*}
\epsfxsize=400pt
\epsfbox[100 400 578 800]{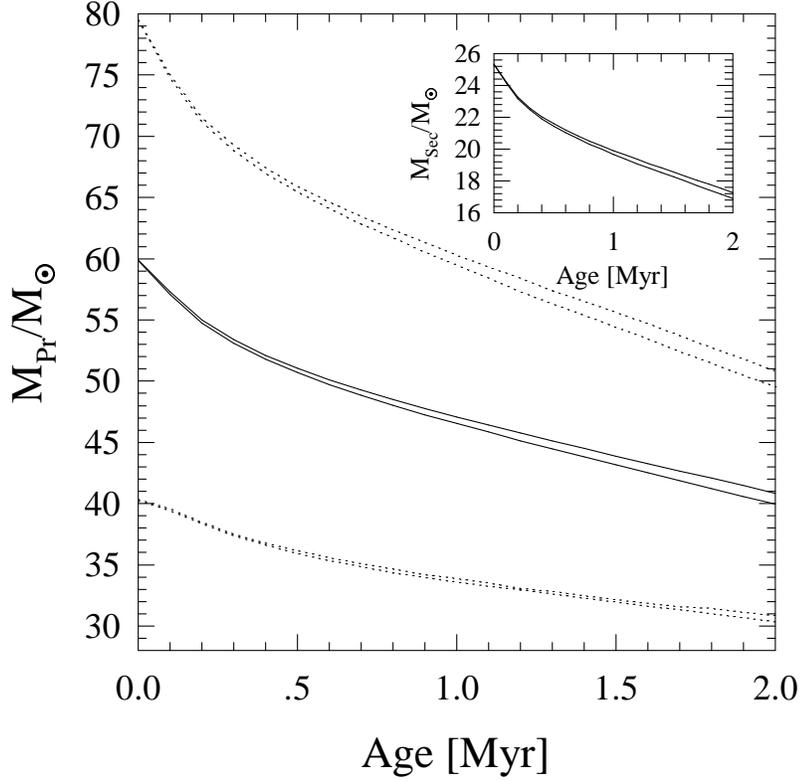} 
\caption{The  mass of  the  components of  HD~93205  deduced from  its
apsidal motion rate as a function  of its assumed age. In each figure,
solid lower  (upper) line corresponds to the  preferred value assuming
$\alpha_{ov}=0.25$  ($\alpha_{ov}=0.4$). Short dashed  lines represent
its $1\sigma$ uncertainty. For more details, see text.
\label{fig:hd93205-m1}}
\end{figure*}

In Fig.~\ref{fig:hd93205-m1}, we present the mass $M_1$ of the primary
component of  HD\,93205 as a function  of the age of  the system.  Two
curves are shown, each of them corresponding to a particular choice of
the  overshooting  parameter  ($\alpha_{ov}=0.25,  \ 0.40$).   Let  us
emphasize that  the amount of  overshooting that actually occurs  is a
rather uncertain  quantity so we consider  it as a  free parameter and
study its  influence on the solution of  Eq.~(\ref{eq:fundam}). As can
be  seen from  Fig.~\ref{fig:hd93205-m1}, for  a given  age,  $M_1$ is
almost  insensitive to  our different  choices of  $\alpha_{ov}$. Both
curves  are almost  overlapped over  the whole  range of  ages, though
differences  tend to  increase  with  age. This  is  not
surprising because for  a given mass the initial  model (a ZAMS model)
is the  same in  both cases so  no initial discrepancy  exists between
them.  As  models evolve  both sequences depart  from one  another and
different internal  mass concentrations slowly arise. In  view of this
insensitivity,   we   shall   concentrate   ourselves  on   the   case
$\alpha_{ov}=0.25$ but  there is no  particular reason to  prefer this
value instead of the higher one.

\begin{figure*}
\epsfxsize=400pt
\epsfbox[100 520 578 800]{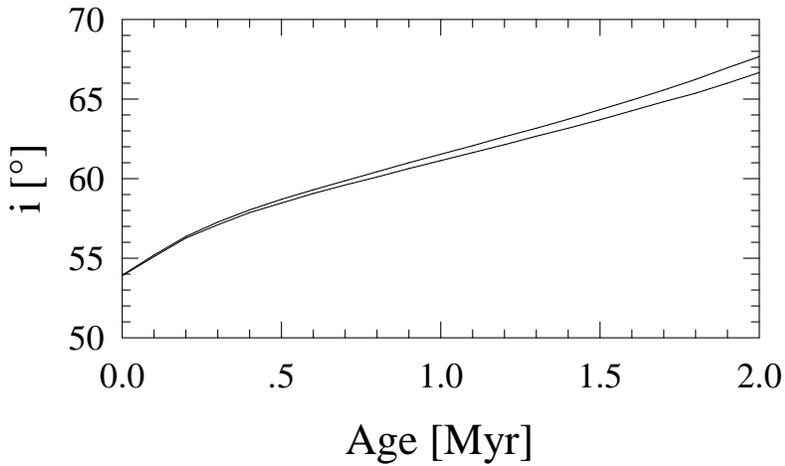} 
\caption{Inclination   of   HD   93205    as   a   function   of   its
age. For more details, see text. \label{fig:hd93205-inc}}
\end{figure*}

Let  us consider  again  Fig.~\ref{fig:hd93205-m1}.  The  mass of  the
primary is  a decreasing function of  the age of the  system.  We find
that its  maximum value, corresponding to  ZAMS models, is  $M_1=60 \pm 19
\rm M_{\odot}$.   This is the  upper limit for  the mass $M_1$  of the
O3\,V component of HD 93205.  At increasing ages, it rapidly decreases
and  reaches  $53  \ \rm  M_{\odot}$  at  just  0.3  Myr and  $46.5 \rm
M_{\odot}$ at  1 Myr  approximately and finally 
$M_1=40 \pm 9 \ \rm M_{\odot}$ at 2 Myr.
Within  observable  quantities,  the  main source  of  uncertainty  in
determining  $M_1$ is  the  apsidal motion  rate  (known up  to a  9\%
accuracy)  and to  a smaller  extent the  projected semi-axis  and the
projected  rotational velocities,  so better  determinations  of these
quantities (especially the apsidal motion rate) are needed in order to
decrease the error  in the determination of $M_1$.  We recall here
that the apsidal motion
rate is a  critical parameter because the necessity  of very long time
baseline (decades) of high-quality observations. Having the mass $M_1$
determined, it is  straightforward to calculate the mass  $M_2$ of the
secondary if we recall (Table \ref{param-hd93205}, see Morrell et al. 2001
for further details)  that the mass ratio $Q=M_2/M_1$  for HD 93205 is
$0.423\pm0.009$.  We find that $M_2$  ranges from $25.3 \pm8\ \rm M_{\odot}$
at the ZAMS down to $17 \pm4  \rm M_{\odot}$ if a rather large value of 2
Myr is  adopted for the  age of the  system.
These mass  values are  in good agreement  with those expected  for an
O8\,V  star like  this  one (consider,  particularly,  the well  known
short-period eclipsing  binary EM~Car,  whose primary component  is an
O8\,V  and its mass  is $22.89  \pm 0.32  \rm M_{\odot}$,  Andersen \&
Clausen  1989).  Once $M_1$  is determined  it is  easy to  obtain the
inclination  $i$ of  the orbit  from Eq.~(\ref{eq:fmasa}).   In 
Fig.~\ref{fig:hd93205-inc} it  is shown the resulting  inclination from the
set of  calculations corresponding  to $\alpha_{ov}=0.25$.  It  can be
seen that the inclination of the system increases as age does. This is
a direct consequence of the behaviour of $M_1$ (which decreases as age
increases) but it is worth noting  that within the whole range of ages
considered the resulting inclination does not allow eclipses to occur.
Indeed, if  we assume that HD  93205 is not  older than 2 Myr  we find
that  $54^{\circ}  \leq  i   \leq  68^{\circ}$,  in  coincidence  with
Antokhina  et  al.   (2000)  who   found  a  most  probably  value  of
$i=60^\circ$.

However,  a problem  arises  when  we try  to  compare the  luminosity
derived  from  the corresponding  models  to  the  observed value  for
HD\,93205. Let us explain  this with an example: if  we consider the 
60 M$_{\odot}$  model, it  predicts,  for zero  age,  a radius  $R_1=10.7$
R$_{\odot}$,   and  a  $\log{T_{\rm   eff}}=4.68$,  resulting   in  a
luminosity,  $\log{L}=5.72$\,L$_{\odot}$.    This  corresponds  to  a
bolometric  magnitude, $M_{\rm  bol}=-9.55$,  which is  almost  one
magnitude fainter than $M_{\rm  bol}=-10.41$ derived by Morrell et al.
(2001) from the visual magnitude  of the O3\,V component of HD\,93205,
the distance modulus of 12.55 obtained by Massey \& Johnson (1993) for
Tr\,16, and  the bolometric  correction (BC) for  an O3\,V  star taken
from the  calibration by Vacca,  Garmany \& Shull (1996).   This large
disagreement between the  expected and observed bolometric magnitudes,
points  to   a  large  error  in   some  (or  any)   of  the  involved
assumptions. If  the distance  modulus is right,  then we  can suspect
that the BC must be wrong  by about one magnitude.  On the other hand,
the  distance  modulus of  the  Carina Nebula  is  still  a matter  of
discussion. Distance modulus of the order of that derived by Massey \&
Johnson (1993) arise from the consideration of  color-magnitude
diagrams for the stellar component of the clusters.   
Some other independent  determinations, like  the recently
obtained  by  Davidson et  al.  (2001)  from  kinematic study  of  the
Homunculus nebula surrounding Eta Car, give distance modulus as low as
11.76,  which would significantly  decrease the  referred discrepancy.
But, if we suppose this last distance modulus to be correct, then all of
the stars in  Tr\,16 will  have $M_{\rm V}$  about 0.8  magnitudes 
fainter than the values accepted to  date.
Here we arrive at a  point whose importance is
obvious for  many astrophysical issues,  and deserves to  be carefully
studied.  The referred discrepancies might also arise in a combination
of different  sources of error  (BCs, distances, and  adopted absolute
magnitude scale for ZAMS stars). A detailed discussion of these issues
will be presented in a forthcoming paper.

Finally, let us comment briefly that tidal contribution to the apse 
motion of HD~93205 is the most important, ranging from about 60\% at the ZAMS
to 70\% at 2~Myr. The rotational contribution ranges from 30\% to 20\% and
the relativistic one is almost constant and approximately 10\% of the
apsidal motion rate is due to this effect. In this  sense, HD  93205 could  
be classified  as a  relativistic binary system (Claret 1997).

\begin{table}
\begin{minipage}{60mm}
\caption{Observed parameters for HD 93205}
\label{param-hd93205}
\begin{tabular}{@{}lll}
\hline
$a_1\  \sin  i$\   [km]  &  $(1.03  \pm  0.02)\,10^7$   &  Morrell  et
al. (2001)\\  $a_2\ \sin i$\  [km] & $(2.44  \pm 0.02)\,10^7$ &  \multicolumn{1}{c}{''} \\
$K_1$\ [km  s$^{-1}$] & $132.6\pm2.0$ &  \multicolumn{1}{c}{''} \\ $K_2$\  [km s$^{-1}$] &
$313.6\pm1.8$ &  \multicolumn{1}{c}{''} \\ $P$\ [days]  & $6.0803\pm0.0004$ & \multicolumn{1}{c}{''}  \\ $e$ &
$0.370\pm0.005$ &  \multicolumn{1}{c}{''} \\ $M_1\,\sin^3 i$ [M$_\odot$]  & $31.5\pm1.1$ &
\multicolumn{1}{c}{''} \\ $M_2\,\sin^3 i$ [M$_\odot$]  & $13.3\pm1.1$ & \multicolumn{1}{c}{''} \\ $Q(M_2/M_1)$
&  $0.423\pm0.009$  &  \multicolumn{1}{c}{''}\\ $\dot{\varpi}$  [$^\circ$\,days$^{-1}$]  &
$0.00533\pm0.00051$  & \multicolumn{1}{c}{''}  \\  $V_1\,\sin  i$ [km  s$^{-1}$]  & 135  &
Howarth et al. (1997) \\ $V_2\,\sin i$ [km s$^{-1}$] & 145: & \multicolumn{1}{c}{''}\\
\hline
\end{tabular}
\end{minipage}
\end{table}

\section{Conclusions} \label{sec:conclu}

We present a method to calculate masses for components of non-eclipsing
binary systems if  their apsidal motion rate is  provided.  The method
consists  in solving  Eq.   (\ref{eq:fundam}) if  the  radius and  the
internal structure constant  of each component can be  obtained from a
grid of stellar evolution calculations.  In order to test this method,
we have  selected some eclipsing  binary systems and have  derived the
masses of their components. A very good agreement was achieved between
masses obtained with our method and those derived from the analysis of
their radial velocity and light curves.

The main  goal of this article,  besides presenting the  method, is to
calculate the masses of the  components of HD\,93205. This is an O3\,V
+ O8\,V  system. Its O3\,V  component has the earliest  known spectral
type of a normal star found in  a double-lined close  binary system,
thus potentially being a very  massive star. Although HD\,93205 is not
an eclipsing binary,  Morrell et al. (2001) have  measured its apsidal
motion rate and found  it to be $\dot{\varpi}=0\fdg0324 \pm 0\fdg0031$
per orbital cycle  so we have been able to  apply the method presented
here to  this system.  The resulting mass  of the primary star ($M_1$) is
obtained as a function of the assumed age of the system.  HD\,93205 is
a highly eccentric system ($e=0.370\pm  0.005$) which suggests a very 
low age. However, we do not adopt a particular value for the age as
its determination is  quite uncertain, and prefer to  consider a range
of ages  starting at the  ZAMS. We find  that for zero age  models the
resulting  mass   is  $M_1=60\pm19  \  \rm  M_{\odot}$   and  that  it
monotonously     decreases    as     age     is    increased     (Fig.
\ref{fig:hd93205-m1}), reaching $M_1=40\pm 9\ \rm M_{\odot}$ at 2 Myr.
Now, if  we take into account  the mass ratio $Q=0.423$  for HD 93025,
the  mass  of  the  secondary  lies in  the  range  $M_2=25.3-17\  \rm
M_{\odot}$ for this range of  ages.  It is worth mentioning again that
these $M_2$ values  are in good agreement with  the masses derived for
other O8~V stars in eclipsing binaries such as the well studied system
EM~Car (Andersen \&  Clausen 1989).  The mass value  derived for $M_1$
is also in  the range (52 -- 60 M$_\odot$)  obtained from the observed
$Q$ assuming a ``normal'' mass for  the O8\,V secondary component (i.e. 
22 -- 25 M$_\odot$).  
In addition, we have estimated the inclination of the
system through  Eq.  (\ref{eq:fmasa})  and the results  obtained (Fig.
\ref{fig:hd93205-inc}) are consistent with the non-eclipsing condition
of HD~93205.

Our results corresponding to zero age  give an upper limit to the mass
of  the O3\,V  component of  HD\,93205, a  result that  places  a strong
constraint to  the masses of  theoretical stellar models for  the most
massive stars.   Also, the luminosity derived from  the stellar models
for the O3\,V component rises a problem when compared with the
observed  value,  being  the  theoretical  $M_{\rm  bol}$  almost  one
magnitude fainter  than the value derived from  the observations. This
discrepancy  raises the  need of  reviewing both  the distance  and BC
scales  for the  earliest  type ZAMS  stars,  a subject  that will  be
addressed in the near future.
 
\section*{Acknowledgements} We acknowledge our anonymous referee for 
comments that helped to improve this work.
RHB acknowledges financial support from 
Fundaci\'on Antorchas (Project No. 13783-5).

\label{lastpage}
\newpage


\newpage

\newpage

\newpage

\newpage

\newpage

\newpage

\newpage

\end{document}